\documentclass[%
 reprint,
 amsmath,amssymb,
 aps,
 prd,
 lengthcheck,%
]{revtex4-1}

\usepackage{graphicx}

\newcommand{\be}{\begin{eqnarray}}
\newcommand{\ee}{\end{eqnarray}}
\newcommand{\nl}{\nonumber\\}
\newcommand{\e}{\mathrm{e}}
\newcommand{\1}{\mathbb{I}}
\newcommand{\tr}[1]{\left\langle #1 \right\rangle}
\newcommand{\trs}[1]{\left\langle #1 \right\rangle^2}
\newcommand{\mh}{{\hat m}}
\newcommand{\cL}{{\cal L}}
\newcommand{\cV}{{\cal V}}
\newcommand{\cS}{{\cal S}}
\newcommand{\cT}{{\cal T}}
\newcommand{\cD}{{\cal D}}
\newcommand{\cZ}{{\cal Z}}
\newcommand{\Zh}{\hat{\cal Z}}

\begin{document}
\title{Staggered chiral random matrix theory}
\author{James C. Osborn}
\affiliation{Leadership Computing Facility, Argonne National Laboratory,
 Argonne, IL 60439, USA}
\begin{abstract}
  We present a random matrix theory (RMT) for the staggered
  lattice QCD Dirac operator.  The staggered RMT is equivalent to the
  zero-momentum limit of the staggered chiral Lagrangian and includes
  all taste breaking terms at their leading order.  This is an extension
  of previous work which only included some of the taste breaking terms.
  We will also present some results for the taste breaking contributions
  to the partition function and the Dirac eigenvalues.
\end{abstract}
\maketitle

\section{Introduction}

Staggered fermions are one of the commonly used ways to simulate
quarks on a lattice due to their relatively low computational expense.
However they are considerably more difficult to deal with theoretically
due to the presence of extra modes.
A single staggered Dirac matrix yields four flavors
of quarks in the continuum limit due to the fermion doubling problem.
These flavors are mixed at finite lattice spacing and are
conventionally referred to as ``tastes'' to distinguish them from
regular quark flavors.  The breaking of the SU(4) taste symmetry can
be reduced by using improved actions, but the taste breaking (TB) is still
observed in current simulations and must be accounted for when
extracting results.

The form of the taste breaking has been worked out as
corrections to the chiral Lagrangian by Lee and Sharpe
\cite{Lee:1999zxa}, with extension to multiple flavors by Aubin and
Bernard \cite{Aubin:2003mg}.  The staggered chiral Lagrangian
includes all terms of $O(a^2)$ ($a$ is the lattice spacing)
that are consistent with the symmetries of staggered fermions.

Here we will construct a complete random matrix theory for the
staggered Dirac matrix that incorporates all terms of $O(a^2)$.
This is an extension of the work in \cite{Osborn:2003dr} where only
the case of zero topological charge was considered, and not all of the
terms found in the chiral Lagrangian could be reproduced in the RMT.
This RMT can be directly related to the staggered chiral Lagrangian
in the zero momentum limit.
We then expect to be able to use this model to study the effects
of TB on low energy quantities such as the
partition function, chiral condensate, and Dirac eigenvalues.
Additionally, since the standard method to deal with the extra quark
modes is by taking the fourth root (or square root) of the fermion
determinant, we expect to be able to study the interactions of the
TB with the ``rooting'' procedure.
In this work we will study the properties of the RMT itself and save
 the comparisons to direct simulations of lattice QCD for later.

\section{Staggered fermions}

On a 4d lattice the unimproved staggered fermion action,
$\bar\psi [D_s+am] \psi$,
with quark mass $m$ is given by
\be
a m \, \bar\psi_x \psi_x + \frac12 \sum_{\mu=1}^4 \eta_{x\mu} \left[
\bar\psi_x U_x^\mu \psi_{x+\hat\mu} -
 \bar\psi_{x+\hat\mu} U_x^{\mu\dagger} \psi_{x}
\right]
\ee
where $U_x^\mu$ is a set of SU($N_c$) matrices representing the gauge
field and $\eta_{x\mu} = (-1)^{\sum_{\nu<\mu}x_\nu}$.
Here, for convenience, we will only consider the case of $N_c=3$.
The RMT given below will also apply for $N_c\ge3$, however for
$N_c=2$ the low eigenmodes are known to be in a different universality
class, described by the Gaussian Symplectic Ensemble \cite{Halasz:1995vd}.
One should be able to extend this model to include $N_c=2$ 
in a similar fashion, but we will not pursue that here.

The massless staggered fermion matrix is
anti-Hermitian and thus has purely imaginary eigenvalues.  It also
possesses a particular form of chiral symmetry
\begin{eqnarray}
\{ \Gamma_5, D_s \} = 0 ~,~~ \Gamma_5 = (-1)^{\sum_{k=1}^4x_k}
\label{symm}
\end{eqnarray}
which causes the eigenvalues to come in positive and negative pairs:
$\pm i \lambda$.
It is well known that the above action contains doubler modes so that
a single staggered fermion matrix actually describes four flavors of
fermions in the continuum limit.
An explicit identification of the continuum fermions was given by
Kluberg-Stern, et al. \cite{KlubergStern:1983dg}.
This was done by transforming the 3 color degrees of
 freedom on the 16 sites of each $2^4$ hypercube into a basis of
 four Dirac fermions (12 components each) on a lattice of half the size
 in each direction.
This transformation is not unique.
In their basis the expansion of the staggered fermion operator
starts as
\be
 (2a)^4 \left\{
  (\gamma_\mu \otimes \1_4) D_\mu + m ( \1_4 \otimes \1_4 )
  -a(\gamma_5 \otimes \xi_{\mu 5}) D_\mu^2 + \ldots
 \right\} . \nonumber\\
\label{ksform}
\ee
The notation $(S\otimes T)$ is used to denote the outer product of a
$4\times 4$ spin matrix $S$ and a $4\times 4$ taste matrix $T$ with
$\xi_\mu = \gamma_\mu^*$, $\xi_{\mu 5} = \xi_\mu \xi_5$,
and $\1_4$ is a $4\times 4$ identity matrix.
The first part is just the standard Dirac operator for four identical
flavors with mass $m$.
The remaining term is suppressed by a factor of the lattice spacing and
breaks the SU(4) taste symmetry.
There is also a term of $O(ag)$, and the remaining
 corrections are at least $O(a^2)$.

In the full theory one usually describes the low energy behavior in
terms of an effective chiral Lagrangian.  For staggered fermions this
has been worked out to order $a^2$ and is given by
 \cite{Lee:1999zxa,Aubin:2003mg}
\be
\label{cV}
\cL = \frac{F^2}{8} \tr{\partial_\mu U \partial_\mu U^\dagger} 
-\frac{1}{2} \Sigma_0 m \tr{U + U^\dagger}
+ a^2 \cV
\ee
where $F$ and $\Sigma_0$ are the low energy constants related to the
pion decay constant (with the convention that the physical value for
 $F \approx 131$ MeV)
and the (absolute value of the) chiral condensate respectively.
Here and everywhere below, $\tr{X}$ will stand for the trace of $X$.
There is also a mass term for the taste singlet pion
 (analogue of the $\eta'$) that we have dropped.

The taste breaking terms can be divided into two parts $\cV =
\cV_{1t} + \cV_{2t}$.  The first part contains the single-trace terms
\be
\label{sL}
-\cV_{1t} &=&
  C_1 \tr{ \xi_5 U \xi_5 U^\dagger } \nl
&+& C_3 \frac{1}{2} \sum_\mu \left[ \tr{ \xi_\mu U \xi_\mu U } + h.c. \right] \nl
&+& C_4 \frac{1}{2} \sum_\mu \left[ \tr{ \xi_{\mu 5} U \xi_{5 \mu} U } + h.c. \right] \nl
&+& C_6 \sum_{\mu<\nu} \tr{ \xi_{\mu\nu} U \xi_{\nu\mu} U^\dagger }
\ee
and the second part has the two-trace terms
\be
\label{c25tr2}
-\cV_{2t} &=& 
    C_{2V} \frac{1}{4} \sum_\mu \left[ \tr{ \xi_\mu U } \tr{ \xi_\mu U } + h.c. \right] \nl
&+& C_{2A} \frac{1}{4} \sum_\mu \left[ \tr{ \xi_{\mu5} U } \tr{ \xi_{5\mu} U } + h.c. \right] \nl
&+& C_{5V} \frac{1}{2} \sum_\mu \left[ \tr{ \xi_\mu U } \tr{ \xi_\mu U^\dagger } \right] \nl
&+& C_{5A} \frac{1}{2} \sum_\mu \left[ \tr{ \xi_{\mu5} U } \tr{ \xi_{5\mu} U^\dagger } \right] .
\ee
Note that in the original Lee-Sharpe Lagrangian, the two-trace terms were
 Fierz transformed into one-trace terms of the form
\be
\label{c25tr1}
-\cV_{25} &=& 
  C_2 \frac{1}{2} \left[ \tr{ U^2 } - \tr{ \xi_5 U \xi_5 U } + h.c. \right] \nl
&+& C_5 \frac{1}{2} \sum_\mu \left[ \tr{ \xi_\mu U \xi_\mu U^\dagger }
                                - \tr{ \xi_{\mu5} U \xi_{5\mu} U^\dagger } \right] .~~~~
\ee
This is valid in the one flavor case, but not when extending to
 multiple flavors \cite{Aubin:2003mg}.
We will see below that the staggered RMT naturally leads to the two-trace form.

\section{Chiral random matrix theory}

The standard partially quenched chiral random matrix theory can be written as
\be
\label{RMTZ}
\cZ^{RMT}_{N_f,N_b} =
\int dW p(W) \frac{\prod_{f=1}^{N_f} \det(\cD_0+m_f)}{\prod_{b=1}^{N_b}
 \det(\cD_0+m_b)}
\ee
where $W$ is a $(N+\nu) \times N$ complex matrix with $\nu$ the absolute
 value of the topological charge and the Dirac operator is represented by
\cite{Shuryak:1992pi}
\be
\cD_0 = \left( \begin{array}{cc}
0 & i W \\
i W^\dagger & 0
\end{array} \right) ~.
\ee
It has been shown that the partition function is universal for a large class
of weights \cite{Nishigaki:1996zz,Akemann:1996vr},
but for convenience here we take the simplest form of a Gaussian,
\be
p(W) = \exp(-\alpha N \tr{W^\dagger W})
\ee
with $\sqrt{\alpha} = \Sigma_0 V/ 2N$ ($V$ is the four volume).
This model is the chiral extension of the Gaussian Unitary Ensemble (GUE).
For a review of chiral random matrix models, see \cite{Verbaarschot:2000dy} and
its references.

By now it is well established that the chiral RMT (including all spectral properties)
is equivalent to the zero-momentum sector of the chiral effective theory
 \cite{Osborn:1998qb,Damgaard:1998xy,Basile:2007ki}.
The equivalence is
established through the partially quenched partition functions
\be
\Zh^{RMT}_{N_f,N_b}(\{\mh_f\},\{\mh_b\}) =
\cZ^{eff(0)}_{N_f,N_b}(\{\mh_f\},\{\mh_b\}) ~.
\ee
Here $\Zh^{RMT}$ is the RMT partition function in the microscopic limit,
defined by
taking the limit $N,V\to\infty$ while keeping $\mh = m V \Sigma_0$ fixed.
The chiral effective theory at zero momentum
 (also called the $\epsilon$-regime \cite{Gasser:1987ah})
contains just a mass term
\be
\cZ^{eff(0)}_{N_f,N_b}(\{\mh_f\},\{\mh_b\}) =
\int dU \det(U)^\nu \e^{\frac12 \tr{\hat M (U + U^\dagger)}} ~~~~
\ee
with $\hat M = \mathrm{diag}(\{\mh_f\},\{\mh_b\})$.
The above expression is a supersymmetric generalization of the
usual fermionic chiral Lagrangian so the determinant and trace must
be taken to their supersymmetric equivalents, and the integration is
now over a supersymmetric manifold that is compact in the fermionic
sector but noncompact in the bosonic sector \cite{Damgaard:1998xy}.

Since the TB terms contain no derivatives, they will also
contribute to the zero-momentum partition function by multiplying
the integrand by the extra factor $\exp(-a^2 V \cV)$.
Below we will establish the equivalence between the partition functions
including TB only for the simpler fermionic case.  In principle,
it is necessary to show this for the partially quenched partition functions
as well, in order to establish an equivalence of valence quantities including
the Dirac eigenvalues.  For now we will assume that this equivalence holds
and that it could be obtained from an extension of the proof for the partition
functions without TB.

\begin{table*}
\begin{centering}
\begin{tabular}{|c|c|c|}
\hline
$\cT$ & $S_\cT$ & $ - V \cV$ \\
\hline
$\left( \begin{array}{cc} 0 & i X \\ i X^\dagger & 0 \\ \end{array} \right) \otimes \Gamma$ &
$\beta N \tr{ X^\dagger X }$ &
$\frac{\alpha N}{\beta} \tr{ \Gamma U \Gamma U^\dagger }$ \\
\hline
$\left( \begin{array}{cc} i A & 0 \\ 0 & i B \\ \end{array} \right) \otimes \Gamma$ &
$\beta N [ \tr{ A^2 } + \tr{ B^2 }]$ &
$\frac{\alpha N}{4 \beta} \tr{ \Gamma U \Gamma U + \Gamma U^\dagger \Gamma U^\dagger }$ \\
\hline
$\left( \begin{array}{cc} i b\otimes\1_{N+\nu} & 0 \\ 0 & i b\otimes\1_{N} \\ \end{array} \right) \otimes \Gamma$ &
$\beta N b^2$ &
$\frac{\alpha N}{4 \beta} \trs{ \Gamma U + \Gamma U^\dagger }$ \\
\hline
$\left( \begin{array}{cc} i c\otimes\1_{N+\nu} & 0 \\ 0 & -i c\otimes\1_{N} \\ \end{array} \right) \otimes \Gamma$ &
$\beta N c^2$ &
$\frac{\alpha N}{4 \beta} \trs{ \Gamma U - \Gamma U^\dagger }$ \\
\hline
\end{tabular}
\end{centering}
\caption{
Mappings from corrections to the chiral RMT [$\cT$ in Eq. (\ref{cT})]
 to corrections to the chiral Lagrangian [$\cV$ in Eq. (\ref{cV})].
The Gaussian weight is given by $\exp(-S_\cT)$.
The first two types of taste breaking terms are similar to the ones
appearing in \cite{Osborn:2003dr}.  The last terms are new and will generate
the two-trace terms.
}
\label{table:map}
\end{table*}

\section{Staggered chiral random matrix theory}

To extend the RMT to include taste breaking, we add a term proportional
to $a$ to a taste diagonal Dirac matrix
\be
\label{cT}
\cD = \cD_0 \otimes \1_4 + a \cT
\ee
where $\cT$ incorporates the taste breaking terms considered below.

In \cite{Osborn:2003dr} we considered only the case of $\nu=0$; furthermore,
the additional terms could only reproduce the single-trace terms.
Here we will consider the extension to $\nu\ne0$
 and will also include the two-trace terms.
Similar work has been done for the Wilson Dirac operator
 \cite{Damgaard:2010cz,Akemann:2010em}.

We start with the dominant term which is typically found to be the $C_4$ term
\cite{Lee:1999zxa}.
For arbitrary $\nu$ we can write it as
\be
\label{t4}
\cT = \sum_{\mu} \left( \begin{array}{cc} A_\mu & 0 \\ 0 & B_\mu \\ \end{array} \right)
\otimes \xi_{\mu 5}
\ee
where $A_\mu$ and $B_\mu$ are Hermitian matrices of size $(N+\nu) \times (N+\nu)$ and 
$N\times N$, respectively.
Note that this also has a chiral and taste structure similar to the leading taste
 breaking term in the expansion in (\ref{ksform}).
If we choose a Gaussian weight function for these matrices of the form
\be
\exp( - \beta N \sum_\mu [ \tr{A_\mu^2} + \tr{B_\mu^2} ] )
\ee
then one can show that the chiral Lagrangian will get a correction term
(see Appendix \ref{app:map} for details),
\be
\label{l4}
-\frac{\alpha N a^2}{4 \beta} \sum_\mu
 \tr{ \xi_{\mu 5} U \xi_{\mu 5} U + \xi_{\mu 5} U^\dagger \xi_{\mu 5} U^\dagger }  ~.
\ee
Upon equating this to the $C_4$ term in the effective Lagrangian
 [and noting that there is a $\xi_{5 \mu}$ in (\ref{sL})], we get
$\beta = \alpha N / 2 V C_4$.
Note that we require that $\beta>0$ for convergence of the integrals.
We could have obtained the opposite sign in (\ref{l4}) if we multiplied (\ref{t4})
by $i$; however, this would make that term Hermitian.
Thus the sign of $C_4$ is determined by the need to have an anti-Hermitian
Dirac operator in (\ref{t4}).
We will discuss this issue more in the context of the two-trace terms.

The $C_3$ term can be handled in a manner similar to $C_4$.
The $C_1$ and $C_6$ terms can be obtained from matrices of the form
\be
\cT = \left( \begin{array}{cc} 0 & i X \\ i X^\dagger & 0 \\ \end{array} \right)
\otimes \Gamma
\ee
where $X$ is a $(N+\nu)\times N$ complex matrix.
The correction to the chiral Lagrangian for this term is given in
 Table \ref{table:map}.

It was pointed out in \cite{Osborn:2003dr} that the terms in (\ref{c25tr1})
 could be obtained from a RMT using terms similar to the ones above,
 but the corresponding RMT would contain Hermitian pieces,
 instead of being strictly anti-Hermitian as is the case of
 the staggered Dirac matrix.
By writing those terms in the two-trace form (\ref{c25tr2}),
 one can now find a way to add them to the RMT while preserving
 the anti-Hermiticity.

To do this we need to make linear combinations of the terms.
For example, we can write the $C_{2V}$ and $C_{5V}$ terms as
\be
\frac{C_V^+}{4} \sum_\mu
\trs{ \xi_\mu (U+U^\dagger) }
+ \frac{C_V^-}{4} \sum_\mu
\trs{ \xi_\mu (U-U^\dagger) } ~~
\ee
with
\be
C_V^{\pm} = (C_{2V} \pm C_{5V})/2 ~.
\ee
We can then linearize each of these terms using a Hubbard-Stratonovich
transformation, such as
\be
\e^{\frac{C_V^+}{4} 
\trs{\xi_\mu (U+U^\dagger)}}
=
\int d\sigma \e^{- \frac{|C_V^+|}{4} \left[ \sigma^2 - 2 \sigma s
\tr{\xi_\mu (U+U^\dagger)} \right]} ~~
\ee
where $\sigma$ is a single real variable
and $s = \sqrt{C_V^+/|C_V^+|}$.
This term now takes the form of a mass term that mixes the tastes
(with a mass matrix $\xi_\mu |C_V^+|\sigma s/\Sigma_0$).
The $C_V^-$ term likewise gives a $\gamma_5$ mass term.
The mappings for these terms from the RMT to the chiral Lagrangian are given
in the last two rows of Table \ref{table:map}.

Note again that we now must have $C_V^\pm<0$ in order for this term to be
anti-Hermitian.  The same condition holds for $C_A^\pm$.
The coefficients $C_{A,V}^-$ are proportional to the ``hairpin'' coefficients
$\delta_{A,V}'$ which appear in one-loop results of
chiral perturbation theory \cite{Aubin:2003mg}.
The other combinations $C_{A,V}^+$ do not appear in one-loop expressions and
therefore have not yet been determined from lattice simulations.
However, the negative sign for $C_{A,V}^-$ is consistent with lattice
measurements \cite{Aubin:2004fs}, as are the positive signs of all
the single-trace coefficients.

The inclusion of the two-trace terms in their current form may seem a
bit {\em ad hoc} since they aren't full matrices like the other terms, but
we can rewrite them in a way that seems more natural.
As an example, we consider a RMT with taste breaking terms of the form
\be
\left( \begin{array}{cc} A + (b + c)\otimes \1_{N+\nu} & 0 \\
 0 & B + (b - c)\otimes \1_{N} \\
 \end{array} \right) \otimes \Gamma
\ee
($A,B$ are Hermitian matrices and $b,c$ are real scalars)
with weight
\be
\exp\left(-\beta N [ \tr{ A^2 } + \tr{ B^2 }]
- \gamma N b^2 - \delta N c^2 \right) ~.
\ee
If we make the substitution $A' = A + b + c$, $B' = B + b - c$, then
 $b$ and $c$ no longer appear as part of the matrix, but appear only
 in the weight.
We can then integrate them out to obtain a new weight function,
\be
\exp\Big(&&
-\beta N \left\{ \tr{ [A' - \bar A' ]^2 } + \tr{ [B' - \bar B' ]^2 } \right\}
\\
&&- \frac{N}{4} \left[ \gamma ( \bar A' + \bar B' )^2
+ \delta ( \bar A' - \bar B' )^2 \right]
[ 1 + O(1/N) ]
\Big) \nonumber
\ee
with $\bar A' = \tr{A'}/(N+\nu)$ and $\bar B' = \tr{B'}/N$.
In this way, we see that the two-trace terms in the chiral Lagrangian are
generated by two-trace terms in the RMT potential.
One could then consider adding other terms such as higher powers of the
matrices in the potential to reproduce higher order terms in the chiral
Lagrangian, though we will not pursue that here.

All the terms of the full staggered RMT (SRMT) are written out in
 Appendix \ref{app:srmt}.
Now that we have the full form of the SRMT, we can examine its
structure more closely.
For this, it is convenient to switch to a basis where the remnant of 
chiral symmetry for staggered fermions (\ref{symm}) is
transformed according to
\be
\left( \begin{array}{cc} \1_{N+\nu} & 0 \\ 0 & -\1_N \end{array} \right)
 \otimes \xi_5 \to \left( \begin{array}{cc}
\1_{4N+2\nu} & 0 \\ 0 & -\1_{4N+2\nu} \end{array} \right) .~
\ee
There are several possible choices of basis, all of which give a
 staggered RMT Dirac operator (at $m=0$) of the form
\be
\left( \begin{array}{cc}
0 & R \\
-R^\dagger & 0 \\
\end{array} \right)
\ee
where $R$ is a $(4N+2\nu) \times (4N+2\nu)$ matrix.
Arbitrarily picking one basis gives an $R$ of the form
(using the terms from Appendix \ref{app:srmt})
\be
\left( \begin{array}{cc}
 [i A_{3\mu} - A_{4\mu} + d_+] \sigma_\mu &
 i W + i X_1 + i X_{6\mu\nu} \sigma_\mu \sigma_\nu^\dagger \\
 i W^\dag -i X_1^\dag -i X_{6\mu\nu}^\dag \sigma_\mu^\dagger \sigma_\nu &
 [i B_{3\mu}+B_{4\mu} + d_-] \sigma_\mu^\dagger    \\
\end{array} \right) ~~~~
\ee
with $\sigma_\mu = (1, -i \vec\sigma^*)$ and
$d_{\pm} = i b_{V\mu}-c_{A\mu} \pm( i c_{V\mu} - b_{A\mu} )$.

Note that since $R$ is a square matrix, in general for nonzero
 lattice spacing there are no exact zero eigenvalues of the RMT.
This agrees with the well-known properties of the
 lattice theory.
Also, one can imagine that if the taste breaking terms are large enough,
 then the detailed structure of $R$ may not matter, and the
 low eigenvalues are described well by a standard chiral RMT at $\nu=0$,
 in agreement with numerical studies
 \cite{BerbenniBitsch:1997tx,Damgaard:1998ie,Gockeler:1998jj,Damgaard:1999bq}.
Below we will take the limit of large taste breaking and
 show that this is indeed the case.

\section{Scales}

For standard staggered chiral perturbation theory (in the $p$-regime)
the size of the taste breaking can be measured by the parameter
 \cite{Aubin:2004fs}
\be
\label{chip}
\chi_{a^2}^{(p)} = \frac{a^2 \bar\Delta}{8\pi^2 F^2}
\ee
where $a^2 \bar\Delta$ is a ``typical'' taste breaking term.
Taking this to be the average pion splitting gives
\be
\bar\Delta = \frac{1}{16}(
 \Delta_P + 4\Delta_V + 6\Delta_T + 4\Delta_A + \Delta_S )
\ee
where the $\Delta_X$ parametrize the mass shift of the pions
above the Goldstone pion (the taste pseudoscalar)
\be
m_{\pi X}^2 = m_{\pi P}^2 + a^2 \Delta_X ~.
\ee
This scale determines the convergence of the taste breaking
parts in the perturbative expansion.

For the zero-momentum chiral Lagrangian ($\epsilon$-regime)
considered here, the relevant parameter is
\be
\label{chie}
\chi_{a^2}^{(\epsilon)} = a^2 V \bar C
\ee
where $\bar C$ is another measure of the strength of the taste
breaking, which we will take to be
\be
\bar C = C_1 + 4 C_3 + 4 C_4 + 6 C_6  = \frac{F^2}{8} \bar\Delta ~.
\ee
We have ignored the contributions from two-trace terms in this
 definition for simplicity, though one could include them if
 needed.
If the scale in (\ref{chie}) is small, then one can calculate quantities
 from a perturbative expansion of the zero-momentum effective theory
 (starting from either the SRMT or the chiral Lagrangian)
 in the taste breaking.
In this case we expect the low eigenvalues to be nearly fourfold
degenerate and form clear ``quartets.''
This has been seen in lattice simulations with improved
 actions \cite{Follana:2004sz,Durr:2004as,Follana:2005km}.

In the opposite limit, $\chi_{a^2}^{(\epsilon)} \gg 1$, the approximate
fourfold degeneracy is strongly broken, and the low eigenvalue spectrum will
resemble that of a single flavor due to the remaining unbroken $U(1)$
Goldstone symmetry of staggered fermions.
We will refer to this limit as strong taste breaking, and the opposite
 limit as weak taste breaking, independent of the value of $\chi_{a^2}^{(p)}$.

In typical lattice simulations, the volume is chosen such that the
lightest dynamical mass stays in the $p$-regime, given by the condition
$m_\pi L \gg 1$ ($m_\pi L \approx 4$ is the usual rule of thumb).
In such simulations the smallest eigenvalues are still typically
described by $\epsilon$-regime calculations since they can be
related to observables with valence quark masses equal to the eigenvalues.
The scale below which eigenvalues can be described by the zero-momentum
 Lagrangian (known as the Thouless energy) in QCD is given by
 \cite{Osborn:1998nf,Osborn:1998nm}
\be
E_c = \frac{F^2}{\Sigma_0 L^2}
\ee
[up to a constant factor of $O(1)$].
It is possible for eigenvalues below this scale to be in the 
strong TB regime ($\chi_{a^2}^{(\epsilon)} \gg 1$),
 while observables at higher scales, e.g. around the dynamical quark mass,
exhibit weak TB ($\chi_{a^2}^{(p)} \ll 1$).
The taste breaking scales are related by
\be
\chi_{a^2}^{(\epsilon)} = \pi^2 V F^4 \chi_{a^2}^{(p)}
 \approx 2 V\, \mathrm{fm}^{-4} \,\chi_{a^2}^{(p)}
\ee
using the physical value for $F$ in the last part.
For lattice simulations with large volumes ($V > 0.5$ fm$^4$)
the $\epsilon$-regime observables will exhibit stronger taste breaking
than the $p$-regime observables.  This is the typical case for simulations where
the dynamical quark masses are kept in the $p$-regime.
For small volumes ($V < 0.5$ fm$^4$) the relationship is reversed.
However, in this case one might find that the dynamical mass is also in
the $\epsilon$-regime so that the parametrization of Eq. (\ref{chip})
does not apply.
Below we will examine the properties of the SRMT in both the weak and
strong TB limits and explore the transition region between the two.

\section{Weak taste breaking}

\subsection{Partition function}
\label{secZ}

Since one copy of the staggered Dirac matrix actually produces
 four tastes, it is common in simulations to take a fractional
 power of the quark determinant to produce the desired number 
 of flavors in the continuum limit.
We now consider the SRMT partition function with optional ``rooting'' given by
\be
\cZ^{SRMT}_{N_q}(\{m\},\{n\}) = 
\int d[\mathcal{D}]
\prod_{k=1}^{N_q}
\mathrm{det}[\mathcal{D}+m_k]^{n_k/4} ~~~
\ee
where the integration measure is over all the Gaussian weights and
the powers $n_k$ can be either positive or negative to produce a partially
quenched theory.

If we expand a determinant to order $a^2$, we get
\be
&&\mathrm{det}[
(\mathcal{D}_0+m_k) \otimes \1_4 +
a \mathcal{T}
]^{n_k/4}
\approx \nl
&&~~~~~~~~~~~~~~~\mathrm{det}[\mathcal{D}_0+m_k]^{n_k}
[ 1 - a^2 (n_k/8)
\langle \mathcal{S}_k^2 \rangle ]
\ee
with
\be
\mathcal{S}_k = 
[(\mathcal{D}_0+m_k)^{-1} \otimes \1_4] 
~ \mathcal{T} ~,
\ee
and we have used the fact that $\tr{\cS_k}=0$.
The $O(a^2)$ partition function is then
\be
\label{zsrmt}
\int d[\mathcal{D}]
\left\{ \prod_{k=1}^{N_q}
\mathrm{det}[\mathcal{D}_0+m_k]^{n_k}
\right\}
\left\{ 1-
\sum_{k=1}^{N_q} \frac{a^2 n_k}{8} 
\langle \cS_k^2 \rangle
\right\} .~~~~~~
\ee
The last term in braces is the correction term for the partition function
due to the TB.  One can easily perform the Gaussian integrations
 over the taste breaking terms in the RMT to obtain the correction factor
\be
1 + \frac{1}{\Sigma_0^2 V^2}
\sum_{k=1}^{N_q}
n_k \Big[
 t_1 \trs{(\mathcal{D}_0+m_k)^{-1}}
+ t_2 \trs{\gamma_5 (\mathcal{D}_0+m_k)^{-1}} \nl
+ t_3 \tr{(\mathcal{D}_0+m_k)^{-2}}
+ t_4 \tr{ [\gamma_5 (\mathcal{D}_0+m_k)]^{-2} }
\Big] ~~~~~
\ee
with the dimensionless coefficients
\be
t_1 &=& a^2 V ( 4 C_3 + 4 C_4 + C_1 + 6 C_6 ) \nonumber\\
t_2 &=& a^2 V ( 4 C_3 + 4 C_4 - C_1 - 6 C_6 ) \nonumber\\
t_3 &=& a^2 V ( C_V^+ + C_A^+ ) \nonumber\\
t_4 &=& a^2 V ( C_V^- + C_A^- ) ~.
\ee
Note that this expression can diverge as $m_k\to 0$ if $\nu n_k =1$.
In general, this expression is not valid at very small masses
since $\cS_k$ can grow large, though for large enough masses
it should be a good approximation.
One could produce an alternate expression that is valid even at $m_k=0$
using eigenvalue perturbation theory.
Its construction will be outlined below.

The correction factor can be further simplified using the identities
\be
\tr{\gamma_5 (\mathcal{D}_0+m_k)^{-1}}
&=& \nu/m_k \nonumber\\
\tr{[\gamma_5 (\mathcal{D}_0+m_k)]^{-2}}
&=& \tr{(\mathcal{D}_0+m_k)^{-1} /m_k} ~.
\ee
One still needs to integrate over the random matrix in $\cD_0$; however,
all quantities can be calculated from known results for the
partially quenched partition functions.

If we write the partition function without taste breaking in (\ref{zsrmt}) as
$\cZ_{N_q}(\{m\},\{n\})$, then
the terms with a single trace can be readily evaluated as derivatives of
 the partition function without taste breaking using the substitutions
\be
\partial_{m_k} \cZ_{N_q}(\{m\},\{n\}) &=&
\int d[\cD_0] \, |D| \, n_k \tr{ (\mathcal{D}_0+m_k)^{-1} } \\
\partial_{m_k}^2 \cZ_{N_q}(\{m\},\{n\}) &=& \\
\int d[\cD_0] |D| \big[ & n_k^2 & \trs{ (\mathcal{D}_0+m_k)^{-1} }
 - n_k \tr{ (\mathcal{D}_0+m_k)^{-2} } \big] \nonumber
\ee
where $|D|$ is short for the product of determinants in (\ref{zsrmt}).
This leaves only the two-trace term to be evaluated.  This requires adding an
 extra quenched pair of quark species to evaluate the extra trace,
\be
 \left. \partial_{m_k} \partial_{m_{f}}
 \cZ_{N_q+2}(\{\{m\},m_{f},m_{b}\},\{\{n\},1,-1\})
 \right|_{m_{f}=m_{b}=m_k} \nl
 = \int d[\cD_0] \, |D| \, n_k \trs{ (\mathcal{D}_0+m_k)^{-1} } ~. ~~~~~~~~~~~~~
\ee

It is straightforward, though somewhat tedious, to evaluate these expressions
in the microscopic limit.
As an example, the one flavor partition function in the microscopic
limit is given by
\be
\hat\cZ_1(\mh) = I_\nu(\mh)
\ee
while the partially quenched $N_f=2, N_b=1$ partition function can
easily be written as a determinant of Bessel functions
\cite{Splittorff:2002eb,Fyodorov:2002wq},
\be
\hat\cZ_{21} =
\frac{
\left| \begin{array}{ccc}
I_\nu(\mh)   &  \mh   I_{\nu+1}(\mh)   & \mh^2   I_{\nu+2}(\mh) \\
I_\nu(\mh_f) &  \mh_f I_{\nu+1}(\mh_f) & \mh_f^2 I_{\nu+2}(\mh_f) \\
K_\nu(\mh_b) & -\mh_b K_{\nu+1}(\mh_b) & \mh_b^2 K_{\nu+2}(\mh_b) \\
\end{array} \right| }{ (\mh_f^2-\mh^2) } ~.~~~~
\ee
From these expressions we get that
\be
\hat\cZ_{1}^{SRMT} \approx I_\nu(\mh) \left(1+ t_2 \frac{\nu^2}{\mh^2} \right) 
+ t_4 \frac{I_{\nu-1}(\mh)+I_{\nu+1}(\mh)}{2\mh} \nl
+ (t_1+t_3) s_1 
- t_3 \frac{I_{\nu-2}(\mh)+2 I_{\nu}(\mh)+I_{\nu+2}(\mh)}{4} ~~~~
\ee
where $s_1$ is the complicated expression
\be
  \frac{\nu^2}{\mh^2} I_{\nu}(\mh) &&
+ \frac{2\nu}{\mh} I_{\nu+1}(\mh)
+ \mh K_{\nu+1}(\mh) I_{\nu+1}^2(\mh) \nl
+ K_{\nu}(\mh)
\Big[&&
\nu I_\nu^2(\mh)
-(3\nu+1) I_{\nu+1}^2(\mh) \nl
&&+\left(\mh-\frac{2\nu(\nu+1)}{\mh}\right) I_{\nu}(\mh) I_{\nu+1}(\mh)
\Big]
\ee
due to the two-trace term.

From this, an expression for the quark mass dependence of the
 chiral condensate in the microscopic limit can be obtained by
\be
\hat\Sigma_1(\mh)/\Sigma_0 = \partial_\mh \ln \hat\cZ_1^{SRMT}(\mh) ~.
\ee
One can obtain expressions for any number of flavors through a similar
procedure.
These formulas would apply to lattice simulations performed in the
$\epsilon$-regime, where $m_\pi L \ll 1$.
For lattice simulations where this doesn't apply, one can instead
consider observables as a function of a valence quark mass that
is in the $\epsilon$-regime.
This will be explored next with the quenched condensate.

\begin{figure*}
 \begin{center}
  \includegraphics[clip,width=\columnwidth]{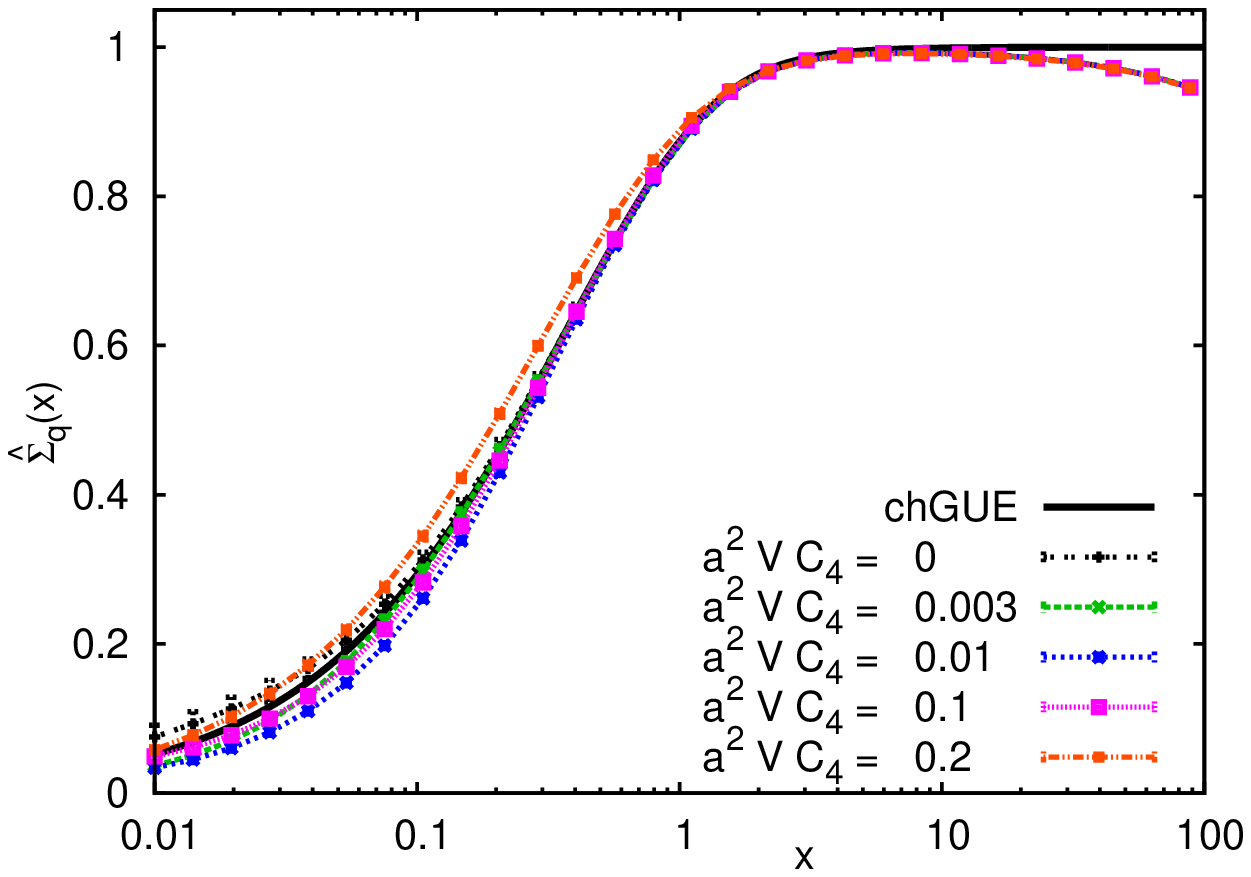}
  \includegraphics[clip,width=\columnwidth]{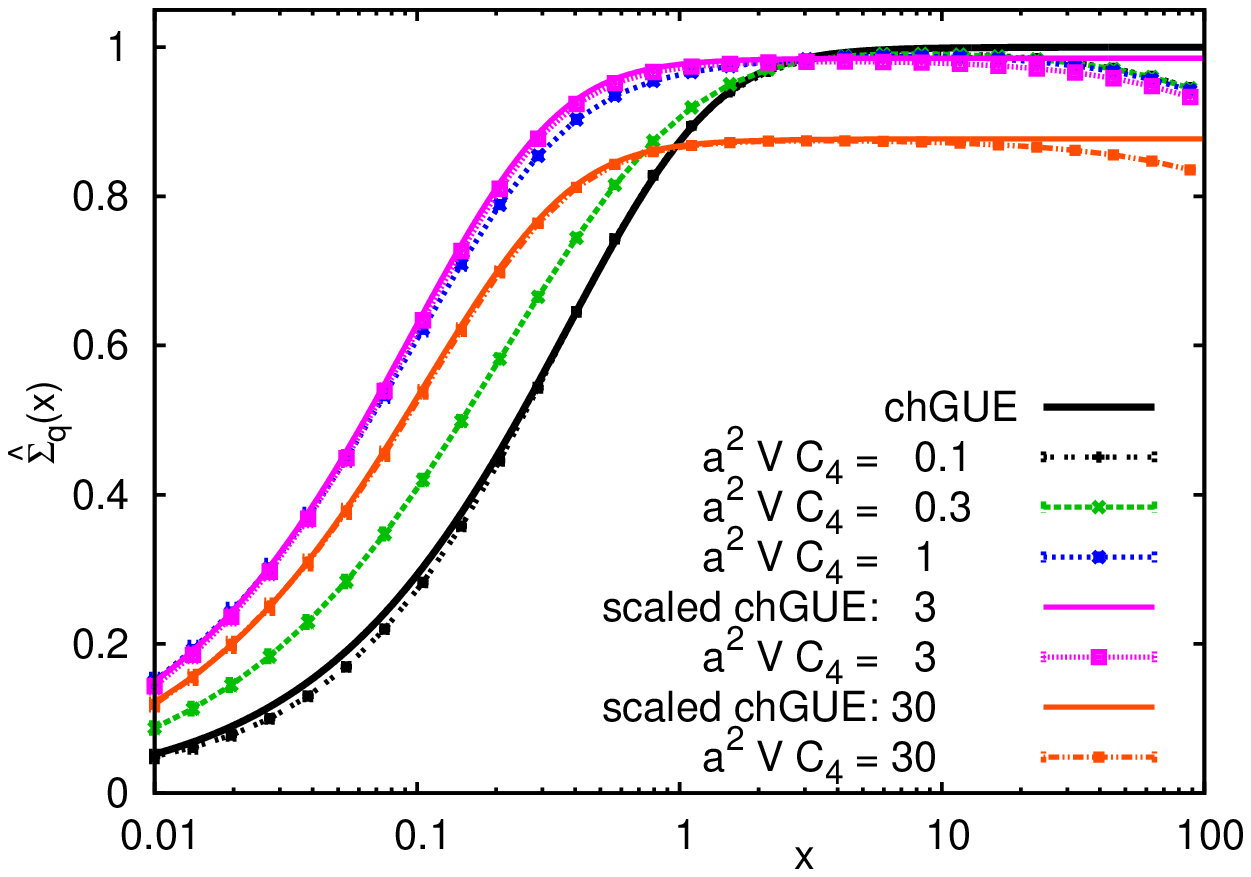}
 \end{center}
 \caption{Quenched chiral condensate for SRMT with $N=400$ and $\nu=0$
 for a range of values of $C_4$ in the weak (left) and strong (right)
 taste breaking regimes.
 The condensate is fairly insensitive to $C_4$ in the weak TB regime.
 In the strong TB regime it agrees well with the chiral GUE result
 (until finite $N$ effects set in at large $x$)
 with a scaled low energy constant $\Sigma_0'$ given in (\ref{stblec}).}
 \label{cc0}
\end{figure*}

\subsection{Quenched condensate}
\label{sec:qc}

One of the simplest valence observables one can look at is the
 quenched condensate.
This can still be useful for comparisons with simulations of full QCD in
 the case that the valence quark mass is much smaller than the dynamical
masses, so that the heavier masses will simply appear quenched compared
to the light valence quark.

The calculation simplifies considerably if we don't try to calculate
the corrections to the quenched partition function first, but instead directly
calculate the corrections to the quenched condensate from the definition
\be
\frac{\hat\Sigma_q(x)}{\Sigma_0} =
 \left. \partial_x \ln \hat\cZ^{SRMT}_2(\{x,y\},\{1,-1\}) \right|_{y=x} ~.~
\ee
Applying this to (\ref{zsrmt}) gives
\be
\cZ_x
&-& t_1 \left( \cZ_{xxy} + \cZ_{xyy} \right)
- t_3 \left( \cZ_{xxx} +2 \cZ_{xxy} + \cZ_{xyy} \right) \nl
&-& 2 t_2 \frac{\nu^2}{x^3}
+ t_4 \left( \frac{\cZ_{xx}+\cZ_{xy}}{x} - \frac{\cZ_{x}}{x^2} \right)
\ee
where $\cZ_x = \partial_x \hat\cZ_{11}(x,y)|_{y=x}$ and similarly
 for higher derivatives.
Using the expression for the microscopic continuum
partition function with $N_f=N_b=1$,
\be
\hat\cZ_{11} =
\left| \begin{array}{cc}
 I_\nu(x) & x I_{\nu+1}(x) \\
-K_\nu(y) & y K_{\nu+1}(y) \\
\end{array} \right| ~,
\ee
we can now evaluate the derivatives.
The result without taste breaking ($\cZ_x$) is \cite{Verbaarschot:1995yi}
\be
\label{qcchgue}
x\, [I_\nu(x)K_\nu(x)+I_{\nu+1}(x)K_{\nu-1}(x)] + \nu/x ~.
\ee
If we write the final expression with TB as $\cZ_x + \sum_k t_k z_k$, then
we have
\be
\label{qcexp}
z_1 &=& z_2 = -2\nu^2/x^3 \\
z_3 &=& z_4 + 2/x -2 K_\nu(x)[I_{\nu+1}(x)+I_{\nu-1}(x)] \\
z_4 &=& -2 K_{\nu-1}(x) I_{\nu+1}(x)/x - 2\nu/x^3 ~.
\ee
Note that the single-trace TB terms enter only through the combination
$t_1 + t_2 \propto C_3 + C_4$, so that $C_1$ and $C_6$ don't contribute while
$C_3$ and $C_4$ only contribute for $\nu \ne 0$.
We will see in numerical simulations below that the quenched condensate
is indeed fairly insensitive to small values of $C_4$ for $\nu=0$.

The (partially) quenched condensate can also be used to calculate the
 eigenvalue density by inverting the Banks-Casher relation \cite{Banks:1979yr}
\be
\Sigma(m) =
 \frac1V \int_{-\infty}^{\infty} \frac{\rho(\lambda)}{m+i\lambda} d\lambda ~.
\ee
The eigenvalue density is obtained from \cite{Osborn:1998qb}
\be
\rho(\lambda) = \lim_{\epsilon\to 0}
 \frac{\Sigma(-i\lambda+\epsilon)-\Sigma(-i\lambda-\epsilon)}{2\pi} ~.
\ee
Care must be taken when evaluating $\Sigma(-i\lambda-\epsilon)$ to
use the correct formula for arguments with negative real parts.
We can avoid this by making use of the fact that the condensate is an
odd function of $m$ (since $\rho$ is even) to get
\be
\rho(\lambda) =
 \frac{\Sigma(i\lambda)+\Sigma(-i\lambda)}{2\pi} ~.
\ee
Applying this to the quenched condensate in the microscopic limit
gives the microscopic quenched eigenvalue density.
The density without taste breaking is \cite{Verbaarschot:1993pm}
\be
\frac{\lambda}{2}\left[
J_\nu(\lambda)^2-J_{\nu+1}(\lambda)J_{\nu-1}(\lambda)\right]
\ee
and the correction term due to taste breaking is
\be
t_3 J_\nu(\lambda)\left[J_{\nu-1}(\lambda)-J_{\nu+1}(\lambda)\right]
-(t_3+t_4) \frac{J_{\nu-1}(\lambda) J_{\nu+1}(\lambda)}{\lambda}. \nl
\ee
Note that the $t_1$ and $t_2$ terms don't contribute at this level.
This expansion will also break down close to zero,
 similar to the condensate.
Expressions that are valid near zero can be obtained from eigenvalue
perturbation theory as discussed in the next section.

\subsection{Eigenvalues}

The full staggered RMT is fairly complicated to work with; however, we
can obtain relatively simple expressions for the effects of taste breaking
 on the splitting of the eigenvalues from eigenvalue perturbation theory.
Since the spectrum without taste breaking contains degeneracies, we must
take these into account.
If $\Psi_k$ is a set of degenerate eigenvectors for the
 eigenvalue $i \lambda_k$,
\be
(\cD_0\otimes \1_4) \Psi_k = \Psi_k i \lambda_k ~,
\ee
then the eigenvalues including the leading order perturbation are given by the
eigenvalues of the matrix
\be
E_k = i \lambda_k + a \Psi_k^\dagger \cT \Psi_k ~.
\ee
For simplicity, we first perform a unitary similarity transformation on $\cD_0$
to put it in the form
\be
U^\dagger \cD_0 U = \left( \begin{array}{ccc}
0 & 0 & 0 \\
0 & 0 & i\Lambda \\
0 & i\Lambda & 0 \\
\end{array} \right) ~.
\ee
with $\Lambda$ a positive diagonal matrix of the nonzero singular values of $W$.
The upper left block of zeros is of size $\nu\times\nu$, representing the zero modes,
while the other two blocks along the diagonal are of size $N\times N$.
The above transformation can be absorbed into the taste breaking terms without
 changing their form, and we will not explicitly write it anymore.
The nonzero eigenvalues of $\cD_0$ written as $\lambda_k$ above are simply
plus or minus the diagonal elements of $\Lambda$.
In this basis the eigenvectors take the form
\be
\Psi = 
\frac{1}{\sqrt{2}}
\left( \begin{array}{ccc}
\sqrt{2}\1_\nu & 0    & 0 \\
0      & \1_N & \1_N \\
0      & \1_N & -\1_N \\
\end{array} \right)
\otimes \1_4 ~.
\ee

We first examine the sector of the $4\nu$ zero modes for which we define $k=0$.
The matrix $E_0$ is just a projection onto the upper left $\nu\times\nu$
 block for all tastes of the taste breaking term.  This gives the form
\be
\label{zmrmt}
i(A_{3\mu} + b_{V\mu} + c_{V\mu})
\otimes \xi_\mu
+ (A_{4\mu} + b_{A\mu} + c_{A\mu})
\otimes \xi_{\mu 5} ~~~~~~~
\ee
with implied summation over $\mu$.
The $A$'s are $\nu \times \nu$ Hermitian matrices and the $b$'s and $c$'s
are real scalars (as given in Appendix \ref{app:srmt}).
The terms are labeled according to their origin in the full SRMT and have the same
Gaussian weights as the corresponding terms in the full SRMT.
Here we see that only the pseudoscalar and tensor terms don't affect the splittings
of the zero modes.
Also, if we ignore the scalars and were to set $C_3=C_4$, then the would-be zero modes
are described by a chiral RMT, for which the eigenvalues are readily found.
However, typically we have that $C_4 \gg C_3$ so that, to a good approximation, we can set $C_3=0$,
which will give a different distribution for the eigenvalues.

We now examine the splitting within a degenerate quartet of eigenvalues.
Here the $4\times 4$ splitting matrix, $\Psi_k^\dagger \cT \Psi_k$, has the form
\be
i d_1 \otimes \xi_5
+ d_{6\mu\nu} \otimes \xi_{\mu \nu}
&+& i x_{V\mu} \otimes \xi_\mu
+ x_{A\mu} \otimes \xi_{\mu 5}
\ee
with $x_{V\mu} = d_{3\mu} + b_{V\mu}$
and $x_{A\mu} = d_{4\mu} + b_{A\mu}$.
All the $d$'s are real scalars; again they are labeled according
 to their origin, and their weight is $\exp(-\alpha N^2 d_X^2/V C_X)$.
Since the $d_{1,3,4,6}$ terms came from matrices in the original SRMT,
 they are different for each quartet.
However, the $b_{A,V}$ terms came from scalars in the SRMT, so these
 variables are in fact the same in all quartets,
 and also in the zero mode sector.
It is these terms that produce the leading order eigenvalue correlations
 between the different quartets and also allow the splittings within each
 quartet to be sensitive to the topological charge in the weak TB regime.
Interestingly these are the terms in staggered chiral perturbation theory
 which don't contribute at one loop.

In Sec. \ref{secZ} we gave an expression for the partition function
that could become singular as $m\to 0$.
Using the eigenvalues from perturbation theory we can derive a similar
expression that treats the zero modes exactly.
Since the $O(a)$ correction to the quark determinant
vanishes, we need to go to second order in eigenvalue perturbation theory.
The expression for the eigenvalues in second order degenerate eigenvalue
 perturbation theory is
\be
E_k = i \lambda_k + a \Psi_k^\dagger \cT \Psi_k - i a^2 \sum_{\ell\ne k}
 \frac{\Psi_k^\dagger \cT \Psi_\ell \Psi_\ell^\dagger \cT \Psi_k}{\lambda_k - \lambda_\ell} ~.
\ee
We can obtain an expression for the quark determinant (and hence the
 partition function) which is accurate to order $a^2$ by multiplying the
 determinants of the $E_k$.
If the determinants for each quartet were expanded in $a$, we would obtain
 an expression for the partition function identical to (\ref{zsrmt}).
However, to produce an expression that is valid at $m=0$,
 the determinants must be handled exactly.
We will not deal with that here.

\begin{figure}
 \begin{center}
  \includegraphics[clip,width=\columnwidth]{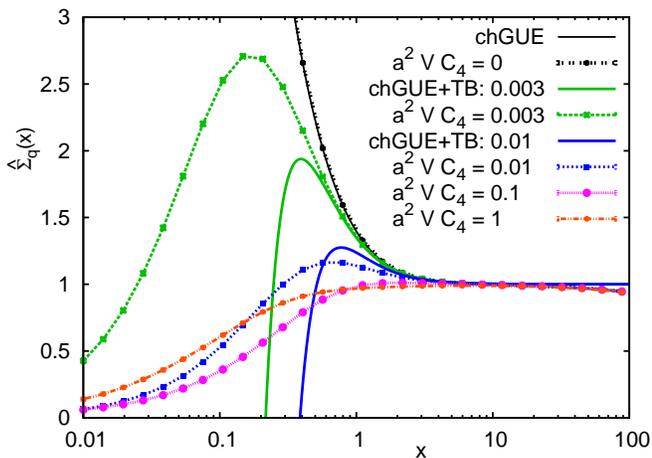}
 \end{center}
 \caption{Quenched chiral condensate for SRMT with $N=400$ and $\nu=1$
 for a range of values of $C_4$ in the weak taste breaking regime.}
 \label{cc1}
\end{figure}

\section{Strong taste breaking}
\label{sec:strongtb}


Earlier we showed how the fermionic SRMT partition function
maps onto the staggered chiral Lagrangian for weak taste breaking.
There we took the large $N$ limit of the SRMT while scaling the terms
$a^2 \alpha/\beta \sim 1/N$.
This means that the $a^2 V C_X$ terms are kept fixed as $N,V \to \infty$.
Now we consider a different limit where the $C_X$ are kept fixed
so that $a^2 \alpha/\beta$ stays constant in the large $N$ limit.
Here the taste breaking must be included in the saddle point equations
when going from the sigma model to the chiral Lagrangian, as mentioned
in Appendix \ref{app:map}.  The inclusion of TB breaks
the Goldstone manifold from SU(4) down to U(1).
Note that if only $C_1\ne 0$ then the symmetry is not fully broken down to U(1),
 but since that is not a likely scenario for common staggered actions, we will 
 not consider it further.

The result for the effective chiral Lagrangian in the strong TB regime
is
\be
\label{cVlt}
\cL = -\frac{1}{2} \Sigma_0' m (U + U^\dagger)
\ee
where $U \in U(1)$.
This has the form of a single fermionic flavor with a modified condensate
given by
\be
\label{stblec}
\Sigma_0' = \frac{4 \Sigma_0}{\sqrt{ 1 + a^2 \bar D }} ~.
\ee
with $a^2 \bar D = a^2 V \bar C /N$.
The factor of 4 is due to the four tastes, and the TB terms
serve to decrease the effective condensate.

Note that this result differs from what we would obtain from the large TB
limit of the weak TB Lagrangian.
That is, if we start with the Lagrangian (\ref{cV}) and take the
 saddle point in the non-Goldstone modes as $a^2 \to \infty$
(or equivalently keeping the $C$'s fixed as $V \to \infty$),
we would get the same form for the effective Lagrangian as in (\ref{cVlt})
but with a condensate that is simply $\Sigma_0' = 4 \Sigma_0 $.
This is because in the weak TB limit $\bar D$ would be scaled to zero,
while it remains finite in the strong TB limit.
Thus, in general, the two limits should be considered distinct even though
some observables may exhibit a smooth transition between them.
We will explore this transition from weak to strong TB
through numerical simulations of the SRMT.

\begin{figure*}
 \begin{center}
  \includegraphics[clip,width=\columnwidth]{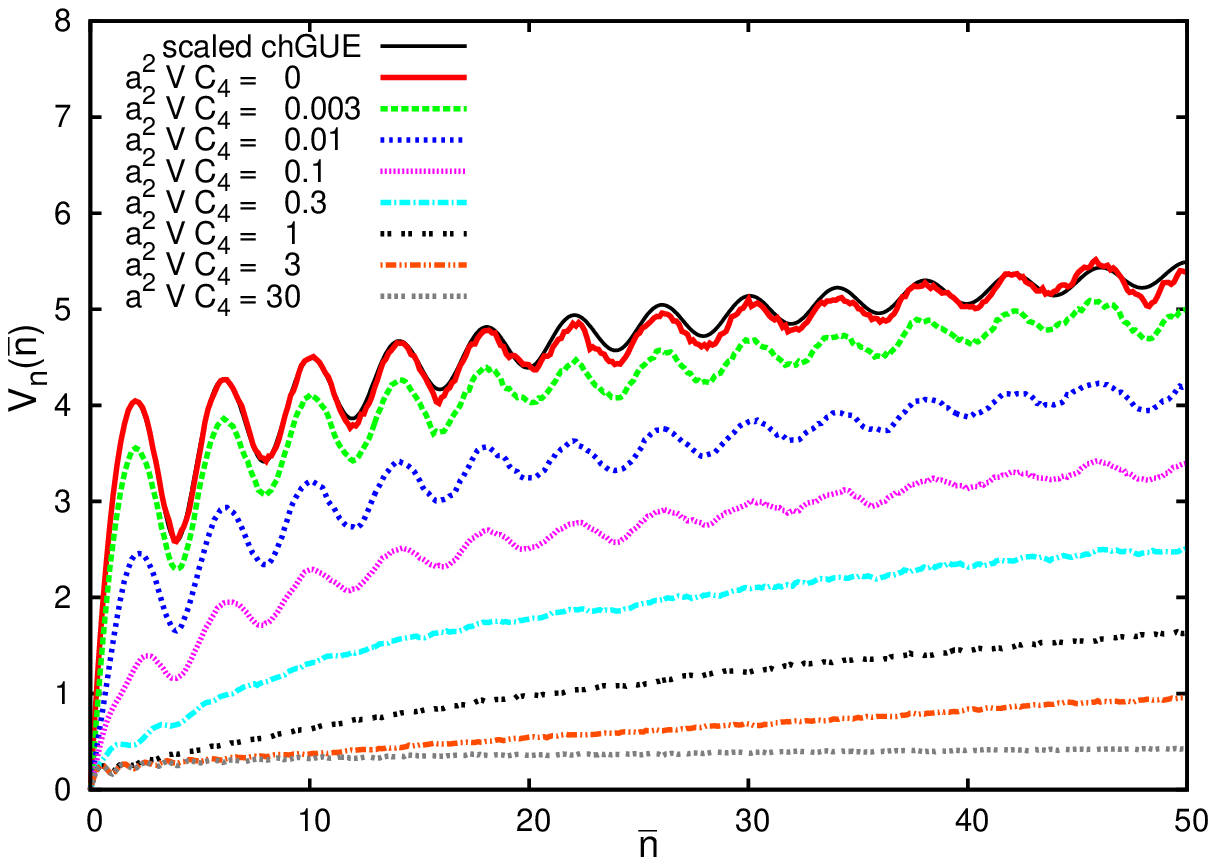}
  \includegraphics[clip,width=\columnwidth]{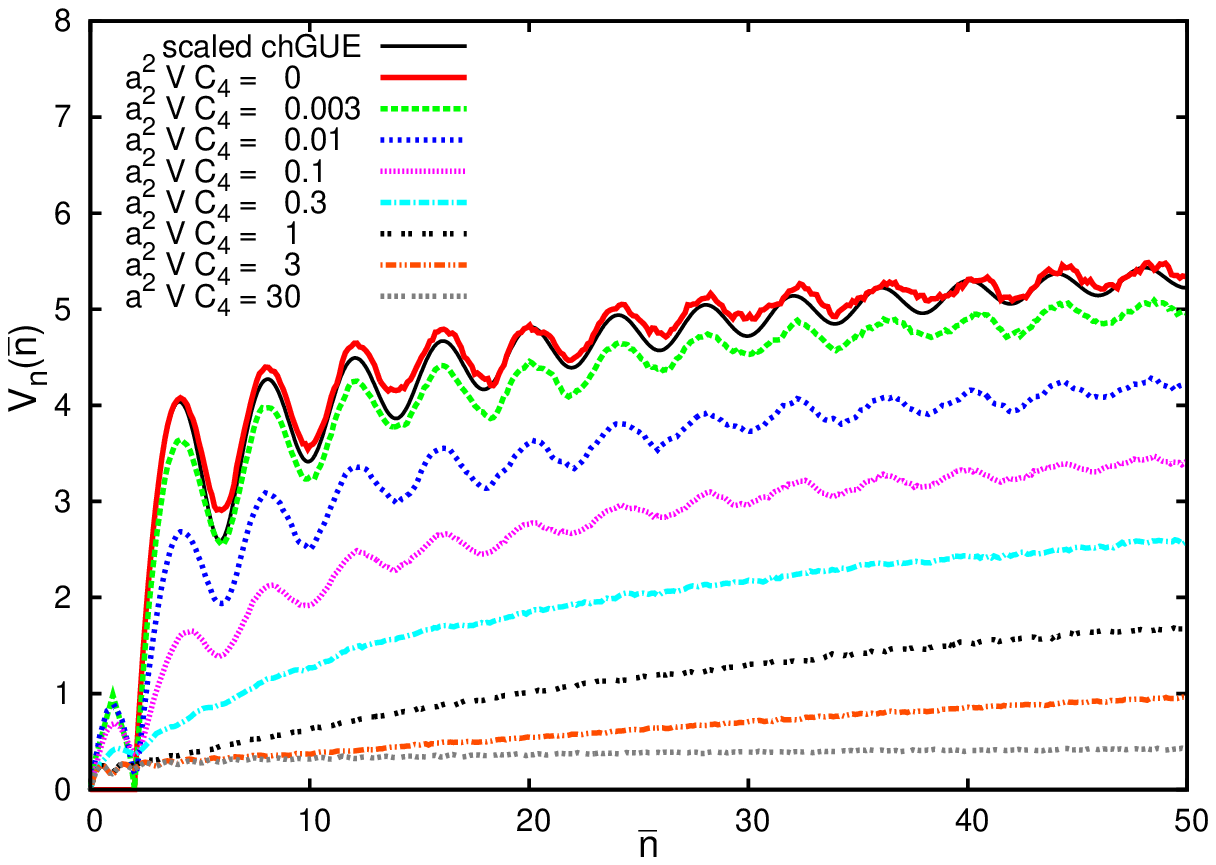}
 \end{center}
 \caption{Number variance for the SRMT with $N=400$ and $\nu=0$, left
 ($\nu=1$, right), for a range of values of $C_4$ showing the
 transition from weak to strong taste breaking.}
 \label{nvall}
\end{figure*}

\section{Numerical results}

Here we present some numerical results obtained from averaging over
10,000 random matrix configurations obtained with $N = 400$
 (and setting $\alpha=1$) at both $\nu=0$ and $\nu=1$.
We varied the value of $c = a^2 V C_4$ from 0 to 100, and all other TB
 coefficients were set to zero.

In Fig. \ref{cc0} we plot the quenched condensate from the SRMT.
In the left panel, the taste breaking parameter $c$ is varied between
$0$ and $0.2$.
For $c=0$ we find the expected agreement with the plotted result from
the chiral GUE (chGUE), Eq. (\ref{qcchgue}), for small $x$.
For larger $x$ we see a discrepancy due to the finite $N$ of the SRMT.
As $N$ is increased, the SRMT results will rise to match the chGUE
curve.
As $c$ is increased, we see some small changes in the quenched condensate for
small $x$, but not for larger $x$.
This is in agreement with our expectations from the expansion (\ref{qcexp})
since we don't expect it to be valid near $x=0$.
At $c \approx 0.2$ we start to see changes in the condensate at larger $x$,
signaling the end of the weak TB approximation.

On the right panel of Fig. \ref{cc0}, we show the quenched condensate as
it leaves the weak TB regime and goes to the strong one.
The scaled chGUE results shown are given by (normalized to a single flavor)
\be
\hat\Sigma_q^{\mathrm{strong}}(x) = \frac{s}{4} \hat\Sigma_q^{\mathrm{chGUE}}(s x)
\ee
with $s = \Sigma_0'/\Sigma_0$ and using Eq. (\ref{stblec}).
For $c>1$ we see very good agreement between the SRMT results and the
above formula, until the finite $N$ effects set in at large $x$.

In Fig. \ref{cc1} we plot the quenched condensate at $\nu=1$
in the weak TB regime.
The chiral GUE result has an explicit $\nu/x$ divergence which is
matched by the $C_4 = 0$ data.
As $C_4$ is increased, the expansion of the condensate gives a TB
correction proportional to $-\nu^2/x^3$ [Eq. (\ref{qcexp})].
This corrected GUE result agrees with the data at $c=0.003$ for $x>0.5$,
while at $c=0.01$ the agreement is found only for approximately $x>2$.
For larger $c$ the expansion breaks down.
At large $c>1$ the $\nu=1$ quenched condensate enters the strong TB regime and
will match the strong TB result of the $\nu=0$ sector.

As previously mentioned, the chiral condensate is related to the eigenvalue
density.  Next we will look at the number variance which is related to the 
two-point eigenvalue correlation function.
It is defined as
\be
V_n(\bar n) = \langle\langle
 \left[ n(\bar n) - \bar n \right]^2 \rangle\rangle
\ee
where $\langle\langle \cdot \rangle\rangle$ denotes the ensemble average.
The function $n(\bar n)$ is the number of eigenvalues between
 $0$ and $\ell$ in a particular configuration with $\ell$
 chosen such that the interval has on average $\bar n$ eigenvalues.

In Fig. \ref{nvall} we plot the number variance for a range of values of
 $C_4$, with all other coefficients set to zero at
both $\nu=0$ and $\nu=1$ topological charges.
At $C_4=0$ the number variance matches that of the chiral GUE
($V_n^{\mathrm{chGUE}}$) \cite{Ma:1997zp}, after the appropriate scaling
\be
V_n^{C_4=0}(\bar n) = 16 V_n^{\mathrm{chGUE}}([\bar n-2\nu]/4)
\ee
due to the fourfold degeneracy of the spectrum and the zero modes.
Note that at $C_4=0$ there are really $4\nu$ exact zero modes; however, we
are considering this to be the limit of $C_4\to 0^+$, where for small
$C_4>0$ the zero modes are split into $2\nu$ positive and negative
near-zero modes.  This gives a shift of the number variance by $2\nu$.

As $C_4$ moves away from zero, we see the number variance drop while retaining
the same pattern of oscillations until $a^2 V C_4 \approx 0.3$.
Here the large oscillations have essentially vanished.
Upon increasing $C_4$ the number variance continues to decrease and a set of
 smaller oscillations appear that coincide with the result from the
 one flavor chiral GUE without any scaling.

\begin{figure*}
 \begin{center}
  \includegraphics[clip,width=\columnwidth]{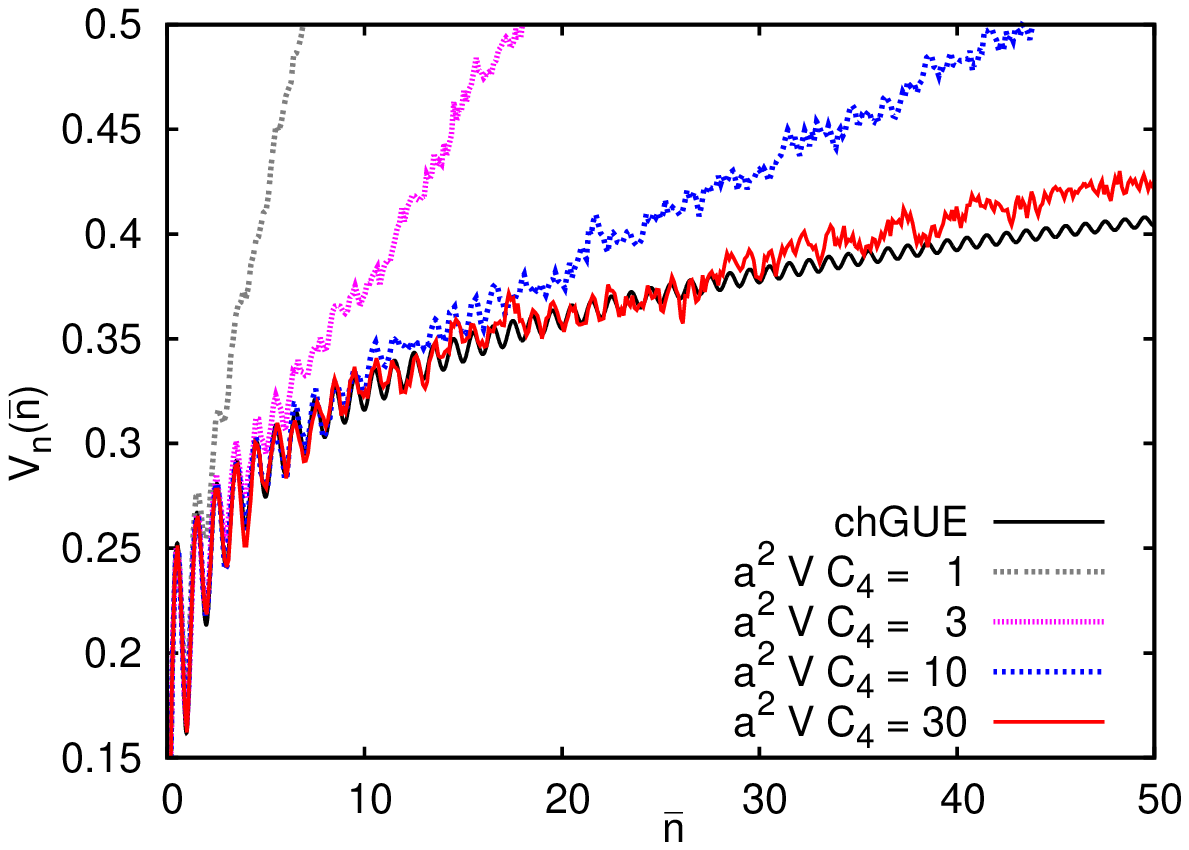}
  \includegraphics[clip,width=\columnwidth]{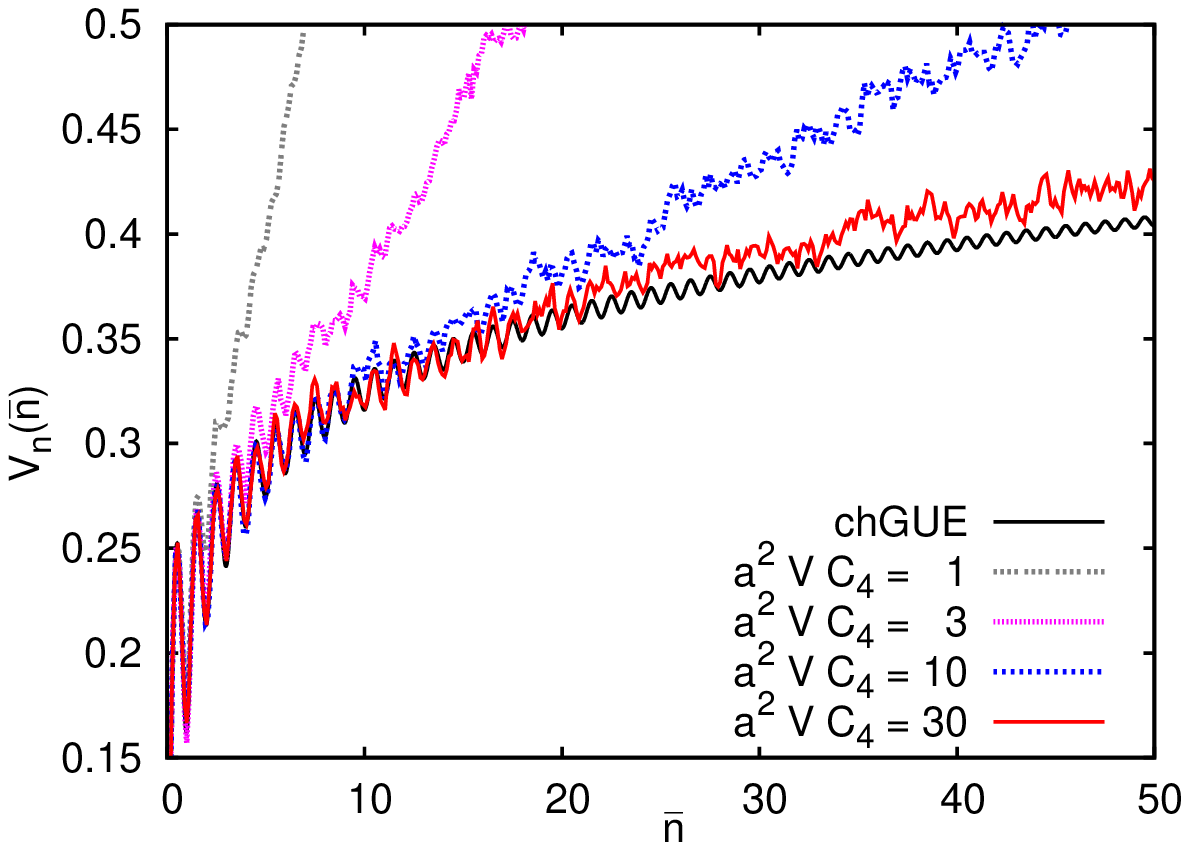}
 \end{center}
 \caption{Number variance for SRMT with $N=400$ and $\nu=0$,
 left ($\nu=1$, right), for a range
 of values of $C_4$ in the strong taste breaking regime.}
 \label{nvstrong}
\end{figure*}

In Fig. \ref{nvstrong} we can see the transition to the strong TB
regime from the same data.
As $C_4$ is increased a larger range of the data falls on top of the 
chiral GUE curve.
The point where the data starts to deviate from the chGUE curve is roughly
equivalent to the value of $a^2 V C_4$.
For example for $a^2 V C_4 = 10$, the number variance agrees with the chGUE
result up to around $\bar n=10$, above which it begins to grow larger than the
chGUE curve.
This is similar to what one sees around the Thouless energy
 \cite{Osborn:1998nm,Osborn:1998nf};
however, in this case it does not signal the breakdown of the RMT, but is
instead the signal of the restoration of an explicitly broken symmetry (taste).
The behavior is independent of the value of $N$.
If one observed a similar behavior of the number variance in lattice
 simulations, this could provide an independent estimate of the size of
 taste breaking.

\section{Summary and future work}

We have presented a completed chiral random matrix theory for staggered fermions
 that includes all the taste breaking effects at order $a^2$.
The SRMT has been shown to be equivalent to the zero-momentum staggered
 chiral Lagrangian in the appropriate limit.
We have also identified a strong taste breaking limit where the SRMT maps
 onto a one flavor effective Lagrangian.

The quenched condensate and number variance evaluated in the SRMT, clearly
 show the transition from the weak to the strong taste breaking regime.
The number variance is particularly interesting since for strong TB, the
 TB scale $a^2 V \bar C$ seems to be directly related to the range of
 agreement to the one flavor chiral GUE result.
We now need to test the results of the SRMT against lattice simulations
 with staggered fermions.

One could also extend the SRMT to include an imaginary quark chemical
 potential.
This can then be used to obtain a measurement of the low energy constant $F$
 \cite{Damgaard:2005ys,Damgaard:2006pu,Damgaard:2006rh,Akemann:2006ru,Akemann2008406}
and could potentially provide more information on the taste breaking.
Similarly one could add a real chemical potential and study the effects of
taste breaking on the complex eigenvalues \cite{Osborn:2004rf,Akemann:2004dr}.
Additionally, one could explore the problems associated with rooting when
the quark determinant is complex \cite{Golterman:2006rw,Svetitsky:2006sn}.

\appendix

\section{Mapping the SRMT to the staggered chiral Lagrangian}
\label{app:map}

The procedure for mapping the chiral RMT onto the zero-momentum chiral
Lagrangian is by now standard \cite{Shuryak:1992pi,Basile:2007ki}.
Here we will consider the case of only fermions,
and will leave the addition of bosonic quarks (ghosts) for later.

We introduce a set of Grassmann variables $\psi^{ts}_i$ and
$\bar\psi^{ts}_i$, with $t$ the taste index, $s=\pm$ the chiral index,
and $i$ the random matrix index.  The determinant can then be written
as
\be
\det(\cD_0 + m + a\cT)
 &=& \int d\bar\psi d\psi \e^{ \bar\psi (\cD_0 + m + a\cT) \psi } \\
 &=& \int d\bar\psi d\psi
 \e^{ m \bar\psi \psi - \tr{(\cD_0+a\cT) \psi \bar\psi} } ~. \nonumber
\ee
In this form the Gaussian integrals over the random matrices can be
readily performed.  This leads to a set of four fermion terms in
the exponential.  For the terms with scalars instead of matrices,
we do not need to perform the Gaussian integrals now, and can wait until
after expanding around the saddle point below.
We will consider the different types of taste breaking terms separately,
and in each case the Gaussian measure is taken to be
\be
\exp(-\beta N \tr{ \cT^\dagger \cT } /4 ) ~.
\ee
The four fermion terms generated by the two types of terms with
 random matrices considered can be summarized as follows:
\be
\label{cterm}
\left(\begin{array}{cc} 0 & i X \\ i X^\dagger & 0 \\ \end{array}\right)
 \otimes \Gamma
&\to&
\frac{a^2}{\beta N}
\tr{\bar\psi^{i+} \psi^{j+} \Gamma_{kj}
 \bar\psi^{k-} \psi^{\ell-} \Gamma_{i\ell}}
\\
\left(\begin{array}{cc} i A & 0 \\ 0 & i B \\ \end{array}\right)
 \otimes \Gamma
&\to&
\frac{a^2}{\beta N} \sum_{s=\pm}
\tr{\bar\psi^{is} \psi^{js} \Gamma_{kj}
 \bar\psi^{ks} \psi^{\ell s} \Gamma_{i\ell}} .~~~~~~
\ee
These can be transformed into fermion bilinears by the Hubbard-Stratonovich
transformation, yielding
\be
\label{rhotb}
&&\exp(-\beta N \tr{\rho^\dagger \rho}
+a \bar\psi^{+} \rho \Gamma \psi^{+}
+a \bar\psi^{-} \rho^\dagger \Gamma \psi^{-}) \\
\label{rhotb2}
&&\exp(-\beta N \tr{\rho_+^2+\rho_-^2}
+a \bar\psi^{+} \rho_+ \Gamma \psi^{+}
+a \bar\psi^{-} \rho_- \Gamma \psi^{-}) ~~~~~~
\ee
respectively, where
$\rho$ is a $4\times 4$ complex matrix and $\rho_\pm$ are $4\times 4$
Hermitian.
The Grassmann integrals can now be performed yielding a determinant.

In the absence of taste breaking, the RMT at this point would be
\be
\int d\sigma \det(\sigma+m)^{N+\nu} \det(\sigma^\dagger+m)^N
\e^{-\alpha N \tr{\sigma^\dagger \sigma}}  ~
\ee
with $\sigma$ a $4\times 4$ complex matrix.
In the large $N$ limit, keeping $\sqrt{\alpha} m N$ fixed (the microscopic limit) one
finds the saddle point solution of $\sigma = U/\sqrt{\alpha}$,
where $U$ is a unitary matrix.
The partition function expanded around this solution then becomes
\be
\int dU \det(U)^\nu \exp(\sqrt{\alpha} m N \tr{U+U^\dagger}) ~.
\ee
Matching to the chiral Lagrangian we get $\sqrt{\alpha} = \Sigma_0 V/2N$.

To include the TB terms we must first choose how to scale those terms with $N$.
If we had scaled the matrix $\rho$ to have the same Gaussian weight as
$\sigma$, then we would get for (\ref{rhotb})
\be
\exp(-\alpha N \tr{\rho^\dagger \rho}
+a\sqrt{\alpha/\beta}[ \bar\psi^{+} \rho \Gamma \psi^{+}
+ \bar\psi^{-} \rho^\dagger \Gamma \psi^{-}]) .~~~~~
\ee
The scaling of the term $a\sqrt{\alpha/\beta}$ determines whether we are
in the weak or strong TB regimes.  If it is held constant then the term with
$\rho$ in the determinant will have a similar magnitude as the term with $\sigma$
and we must include it in the saddle point equations.

For this strong TB regime, at the saddle point we find, for the term given
in (\ref{rhotb}),
\be
\rho = (a/\beta) \Gamma \sigma^{\dagger-1}
\ee
and in (\ref{rhotb2}),
\be
\rho_+ &=& (a/2\beta) \Gamma \sigma^{-1} \\
\rho_- &=& (a/2\beta) \Gamma \sigma^{\dagger-1} ~.
\ee
The two-trace TB terms do not contribute to the saddle point.
The saddle point solution for $\sigma$ has the form
\be
\sigma = c \exp(i \xi_5 \theta)
\ee
for some $c$.
The resulting effective Lagrangian for the
strong TB limit of the SRMT is given in (\ref{cVlt})
with $U = \exp(i\theta)$.

For weak taste breaking we can take $a\sqrt{\alpha/\beta} \sim 1/\sqrt{N}$.
Then the saddle point solution is unchanged by taste breaking and the TB
terms can be expanded around the saddle point solution.
Then the remaining Gaussian integrals can easily be performed,
yielding the set of terms given in Table \ref{table:map}.

\section{SRMT}
\label{app:srmt}

For completeness, we will explicitly write down all the taste breaking terms
from the full staggered RMT.
The SRMT Dirac matrix is given by the form in Eq. (\ref{cT}).
The taste breaking terms $\cT$ are given by the sum of the following terms:
\be
\cT_1 &=& \left( \begin{array}{cc}
0 & i X_1 \\ i X_1^\dagger & 0
\end{array} \right) \otimes \xi_5 \\
\cT_3 &=& \sum_\mu \left( \begin{array}{cc}
i A_{3\mu} & 0 \\ 0 & i B_{3\mu}
\end{array} \right) \otimes \xi_\mu \\
\cT_4 &=& \sum_\mu \left( \begin{array}{cc}
A_{4\mu} & 0 \\ 0 & B_{4\mu}
\end{array} \right) \otimes \xi_{\mu 5} \\
\cT_6 &=& \sum_{\mu<\nu} \left( \begin{array}{cc}
0 & X_{6\mu\nu} \\ X_{6\mu\nu}^\dagger & 0
\end{array} \right) \otimes \xi_{\mu\nu} \\
\cT_V^+ &=& \sum_\mu \left( \begin{array}{cc}
i b_{V\mu} & 0 \\ 0 & i b_{V\mu}
\end{array} \right) \otimes \xi_\mu \\
\cT_V^- &=& \sum_\mu \left( \begin{array}{cc}
i c_{V\mu} & 0 \\ 0 & - i c_{V\mu}
\end{array} \right) \otimes \xi_\mu \\
\cT_A^+ &=& \sum_\mu \left( \begin{array}{cc}
b_{A\mu} & 0 \\ 0 & b_{A\mu}
\end{array} \right) \otimes \xi_{\mu 5} \\
\cT_A^- &=& \sum_\mu \left( \begin{array}{cc}
c_{A\mu} & 0 \\ 0 & - c_{A\mu}
\end{array} \right) \otimes \xi_{\mu 5}
\ee
The Gaussian weights for the individual terms are
\be
\exp&\Bigg(& \frac{\alpha N^2}{8 V} \left[ \tr{
\frac{\cT_1^2}{C_1} + \frac{\cT_3^2}{C_3} + \frac{\cT_4^2}{C_4} + \frac{\cT_6^2}{C_6}
}\right] \nl
&-& \frac{\alpha N^2}{V} \sum_\mu \left[ 
\frac{b_{V\mu}^2}{C_V^+} + \frac{c_{V\mu}^2}{C_V^-} + \frac{b_{A\mu}^2}{C_A^+} + \frac{c_{A\mu}^2}{C_A^-}
\right] \Bigg) ~.
\ee

\bibliography{./refs}

\end{document}